\documentstyle[epsfig]{aprim10}
\input{epsf}

\newif\ifAMStwofonts

\ifoldfss
  \ifCUPmtlplainloaded \else
    \NewTextAlphabet{textbfit} {cmbxti10} {}
    \NewTextAlphabet{textbfss} {cmssbx10} {}
    \NewMathAlphabet{mathbfit} {cmbxti10} {} 
    \NewMathAlphabet{mathbfss} {cmssbx10} {} 
  \fi
  \ifAMStwofonts
    \ifCUPmtlplainloaded \else
      \NewSymbolFont{upmath} {eurm10}
      \NewSymbolFont{AMSa} {msam10}
      \NewMathSymbol{\upi}     {0}{upmath}{19}
      \NewMathSymbol{\umu}     {0}{upmath}{16}
      \NewMathSymbol{\upartial}{0}{upmath}{40}
      \NewMathSymbol{\leqslant}{3}{AMSa}{36}
      \NewMathSymbol{\geqslant}{3}{AMSa}{3E}

    \fi
  \fi
\fi 

\ifnfssone
  \newmathalphabet{\mathit}
  \addtoversion{normal}{\mathit}{cmr}{m}{it}
  \addtoversion{bold}{\mathit}{cmr}{bx}{it}
  \newmathalphabet{\mathbfit} 
  \addtoversion{normal}{\mathbfit}{cmr}{bx}{it}
  \addtoversion{bold}{\mathbfit}{cmr}{bx}{it}
  \newmathalphabet{\mathbfss} 
  \addtoversion{normal}{\mathbfss}{cmss}{bx}{n}
  \addtoversion{bold}{\mathbfss}{cmss}{bx}{n}
  \ifAMStwofonts
    \ifCUPmtlplainloaded \else
      \UseAMStwoboldmath
      \makeatletter
      \new@mathgroup\upmath@group
      \define@mathgroup\mv@normal\upmath@group{eur}{m}{n}
      \define@mathgroup\mv@bold\upmath@group{eur}{b}{n}
      \edef\UPM{\hexnumber\upmath@group}
      \new@mathgroup\amsa@group
      \define@mathgroup\mv@normal\amsa@group{msa}{m}{n}
      \define@mathgroup\mv@bold\amsa@group{msa}{m}{n}
      \edef\AMSa{\hexnumber\amsa@group}
      \makeatother
      \mathchardef\upi="0\UPM19
      \mathchardef\umu="0\UPM16
      \mathchardef\upartial="0\UPM40
      \mathchardef\leqslant="3\AMSa36
      \mathchardef\geqslant="3\AMSa3E
    \fi
  \fi
\fi 

\ifnfsstwo
  \DeclareMathAlphabet{\mathbfit}{OT1}{cmr}{bx}{it}
  \SetMathAlphabet\mathbfit{bold}{OT1}{cmr}{bx}{it}
  \DeclareMathAlphabet{\mathbfss}{OT1}{cmss}{bx}{n}
  \SetMathAlphabet\mathbfss{bold}{OT1}{cmss}{bx}{n}
  \ifAMStwofonts
    \ifCUPmtlplainloaded \else
      \DeclareSymbolFont{UPM}{U}{eur}{m}{n}
      \SetSymbolFont{UPM}{bold}{U}{eur}{b}{n}
      \DeclareSymbolFont{AMSa}{U}{msa}{m}{n}
      \DeclareMathSymbol{\upi}{0}{UPM}{"19}
      \DeclareMathSymbol{\umu}{0}{UPM}{"16}
      \DeclareMathSymbol{\upartial}{0}{UPM}{"40}
      \DeclareMathSymbol{\leqslant}{3}{AMSa}{"36}
      \DeclareMathSymbol{\geqslant}{3}{AMSa}{"3E}
    \fi
  \fi
\fi 

\ifCUPmtlplainloaded \else
  \ifAMStwofonts \else 
    \def\upi{\pi}
    \def\umu{\mu}
    \def\upartial{\partial}
  \fi
\fi

\title[Application of colours in stellar population studies]{Using colours to determine the stellar ages and metallicities of
distant galaxies}

\author[Zhongmu Li]
       {Zhongmu Li\\
       Collage of Physics and electrical information, Dali University, Dali, 671003, China}
\date{}

\pagerange{\pageref{firstpage}--\pageref{lastpage}}
\pubyear{2008}

\begin{document}

\maketitle

\label{firstpage}

\begin{abstract}
The determination of stellar populations of galaxies are important
for studying the formation and evolution of galaxies, because all
galaxies contain many stars and they evolve with galaxies. Spectra
data are usually used to determine the stellar populations of nearby
galaxies as they have the ability to disentangle the degeneracy
between stellar age and metallicity. However, it is difficult to
give similar studies to distant (e.g., $z >$ 0.3) galaxies because
of the lack of reliable spectra data. This is actually limited by
current observational equipments and methods. In fact, the
information of the stellar ages and metallicities of distant
galaxies are crucial for solving the problem of galaxy formation and
evolution. Colours can give us some information of the stellar
populations of distant galaxies. In the paper, I introduce our works
about using colours to estimate the ages and metallicities of the
stellar populations of galaxies. Some potential colours for studying
stellar-population parameters (age and metallicity) and their
sensitivities to stellar-population parameters are reviewed. A new
composite stellar population model that includes both single stars
and binary stars is also introduced.
\end{abstract}


\section{Introduction}

The formation of galaxies is one of the biggest challenges in
astronomy and astrophysics. It is the golden era to solve this
problem, because we have had good background of the formation and
evolution of stars and the universe. In fact, many works have been
done and some progress has been presented. However, it is far from
well understanding galaxies and further researches are needed.
Evolutionary stellar population synthesis is a widely used technique
for studying galaxies. It can give reliable studies to galaxies via
their stellar contents, because all galaxies contain a great deal of
stars and stars contribute mainly to the light of galaxies. A lot of
works studied galaxies using their stellar populations
\cite{Li:2006}. However, most previous works studied the stellar
populations of nearby galaxies, via spectra-like methods, while
optical colours are thought to be unusable for determining two
stellar-population parameters (age and metallicity). This is very
the well-known stellar age--metallicity degeneracy. However, some
studies also showed that colours in different bands are sensitive to
different stellar-population parameters and can possibly be used to
give estimates to stellar populations of distant galaxies, and then
investigations to galaxy formation and evolution. I will review our
works on trying to use colours to study the stellar-population
parameters of galaxies. This will be useful for future studies,
because colours can be obtained much more easily than the spectra of
galaxies.

The structure of the paper is as follows. In section 2, I introduce
the sensitivities of colours to stellar-population parameters and
some potential colours for stellar population studies. In section 3,
I summary a few points that should be considered when using colours
to study stellar populations. In section 4, I introduce a new
stellar population model for using colours to study stellar
populations of galaxies. Finally, in section 5, I give a short
conclusion.

\section{Sensitivities of colours}\label{sec:sens}
Spectra line strength indices can be used to disentangle the stellar
age--metallicity degeneracy because they have different
sensitivities to the age and metallicity of stellar populations.
However, optical colours are thought to be useless for disentangling
stellar age--degeneracy because they have similar sensitivities to
stellar-population parameters. In order to investigate the
possibility of using colours in different bands to study the ages
and metallicities of stellar populations, we investigated the
sensitivities of colours to stellar-population parameters
\cite{Li:2007}. In that work, a simple stellar population model
\cite{Bruzual:2003} and a relative sensitivity method were used. The
results showed that colours in different bands have various
sensitivities to the inputs of stellar populations. In detail, some
colours related to optical bands, e.g., $(B-V)$, $(U-R)$, $(R-I)$
and $(V-I)$, are more sensitive to stellar age, while some other
ones related to near-infrared bands, e.g., $(B-K)$, $(R-K)$, $(V-K)$
and $(I-K)$, to stellar metallicity. However, it showed that every
colour is affected by both stellar age and metallicity. This
suggests that it is impossible to determine stellar age or
metallicity using one colour index. However, using a pair of colours
that consist of an age-sensitive colour and a metallicity-sensitive
colour, the stellar ages and metallicities of galaxies can be
determined. One can refer to our paper \cite{Li:2007} for more
details. When we tried to study the sensitivities of some colours
(hereafter composite colours) including magnitudes on different
photometry systems, we found that some composite colours [e.g.,
$(r-K)$, $(u-K)$, $(u-R)$ and $(i-I)$] have good sensitivities to
stellar-population parameters. Thus they can be used for stellar
population studies. If taken the usual observational errors for
magnitudes, the abilities of different pairs of colours for
disentangling the well-known stellar age--metallicity degeneracy can
be valued \cite{Li:2008a}. It is shown that pairs such as [$(r-K)$,
$(u-R)$] and [$(r-K)$, $(u-r)$] are better for usual stellar
population studies. However, the uncertainties in final stellar ages
and metallicities are somewhat large (near 100\%). This results from
the large observational uncertainties in colours. Because different
surveys have different observational uncertainties, the errors in
the results of various surveys can be different a lot. In future
surveys, because colour uncertainties will be possibly reduced, it
will be possible to give more accurate constraints on stellar ages
and metallicities of galaxies via colours.

\section{Points should be noted}\label{sec:points}
Colours related to different bands have various sensitivities to
stellar-population parameters and can help us to determine the
stellar ages and metallicities of galaxies. However, some points
should be noted, because simple stellar population models are
usually used, but colours can be affected by, e.g., dust, young
stars, and binary stars in galaxies, and observed colours are
related to the distances of galaxies. In the following part, I will
mention a few points.

\subsection{Effects of young stars}
Because early-type galaxies were thought to have some homogeneous
and old stellar populations, some simple stellar population models
were widely used in the studies of the stellar populations of
early-type galaxies. However, more and more observations showed that
there are recent star formations in those galaxies. It suggests that
young stars are common in all type of galaxies and there should be
more than one stellar populations in a galaxy. In this case, it is
necessary to consider the effects of young stars on the
determination of stellar ages and metallicities of galaxies, as
young stars are usually bright and can contribute much to the light.
One \cite{Li:2007b} of our works studied how young stars can affect
the stellar-population parameters determined by colours. That work
shows that if there were two stellar populations (an old and a young
one with the same metallicity) in a galaxy, the younger the age or
the larger the mass fraction of the young component, the bluer the
colours of the galaxy. When one gives estimates to stellar ages and
metallicities via comparing a pair of colours of galaxies to those
of theoretical simple stellar populations, younger ages and richer
metallicities are usually obtained. Therefore, one should take the
effects of young stars into account when studying stellar-population
parameters using colours.

\subsection{Effects of binary interactions}
Because it is easier to model stellar populations via single stars,
most widely used stellar population models are single-star stellar
population models (ssSSPs), which do not take the effects of binary
interactions into account. However, binary stars are common and they
evolve differently from single stars. Two of our works
\cite{Li:2008b,Li:2008c} show that binary interactions make stellar
populations less luminous, while making colours bluer, age-sensitive
line strength indices larger and metallicity-sensitive indices less
compared to ssSSPs. When using colours to determine the
stellar-population parameters of galaxies, different stellar ages
and metallicities will be obtained via ssSSPs and bsSSPs. Usually,
poorer metallicities and similar ages will be given by ssSSP models,
compared to the results obtained via bsSSP models. Therefore, when
investigating the stellar metallicities of galaxies, the effects of
binary interactions should be taken into account. In addition,
although ssSSP and bsSSP models can give similar ages for galaxies,
it is actually much more complicated in practical works, because the
effects of binary interactions are degenerate with
stellar-population mixing. This needs further investigations.

\subsection{Effects of dust and corrections}
Most stellar population models do not take the effects of dust into
account, but dust exists in galaxies and can change the colours of
galaxies. Therefore, the effects of dust should be considered. Many
works about the dust of galaxies have been done, but there is a long
way to go. Although it is difficult to give accurate corrections for
the dust in galaxies, we can reduce the effects of dust in final
results via defining our galaxy samples carefully. If we study
luminous and relatively blue early-type galaxies, dust will affect
our results much more slightly, because there is less dust in such
galaxies. Moreover, there are larger uncertainties in the results of
of galaxies with large (e.g., $>$ 1) red shifts, because the
corrections of the colours of such galaxies have much uncertainties
and they can lead to large uncertainties in final results. However,
this will possibly be improved in the future, following the
development of telescopes and the process of observational data.

\section{A new model for stellar population studies}\label{sec:model}

Although there are many available stellar population models and some
of them are widely used, there are some limitations in those models.
It is necessary to build some new and more advanced models for
stellar population studies and galaxy studies. I introduce a new
model for studying stellar populations of galaxies via colours or
low resolution spectral energy distributions (SEDs). This is a model
that takes both the effects of binary interactions and population
mixing into account. Because binary insteraction and population
mixing are common in galaxies, the theoretical populations of the
new model are closer to those of galaxies.
\begin{figure*} 
  \includegraphics[angle=-90,width=0.8\textwidth]{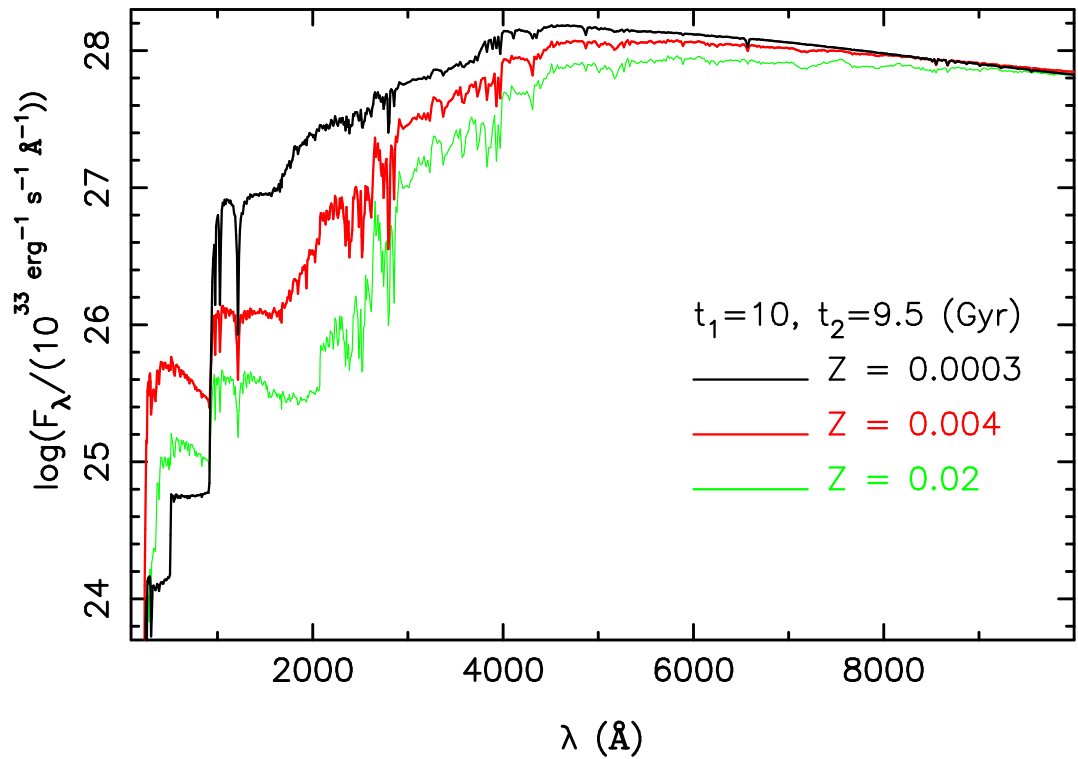}
  \caption{The spectral energy distributions of three composite stellar populations with different metallicities. A fraction of 50\% is taken for binaries.}
\end{figure*}

The model is built using an isochrone database \cite{Li:2008d} of
both single-star and binary-star stellar populations, in which the
evolution of stars was calculated via the rapid stellar evolution
code of Hurley et al. (2002). When building the new model, we used
BaSeL 3.1 spectra library \cite{Westera:2002} to transform the
results of stellar evolution into the SEDs of stellar populations.
Although a galaxy may contains many populations, we build our
population via two stellar populations (an old and a young one).
This makes it easier to calculate the model and possible to estimate
the characteristics of the main components of galaxies. Following
the results of Thomas et al. (2005), a exponentially declining law
with age is taken for the mass fractions of young components in our
work. In such a case, when the model is used for studies, some
mass-weighted stellar ages and metallicities will be obtained for
the components of galaxies. As the main results, the SEDs and
colours of composite stellar populations are calculated. In Fig. 1,
the SEDs of a few populations are shown as an example. We see that
the UV-upturns observed in elliptical galaxies are reproduced by our
new model. In fact, UV SEDs are sensitive to the recent star
formations of galaxies. When we try to study the sensitivities of
colours to the inputs of stellar populations, photometries in
different bands are found to be sensitive to different model inputs.
This will be useful for studying the metallicities and the ages of
the components of the stellar populations of galaxies, and then the
star formation histories of galaxies.

\section{Conclusion and discussion}\label{sec:discuss}

Colours can help us to explore the stellar populations of galaxies,
because different colours have various sensitivities to the inputs
of stellar populations. If the galaxy sample is well selected and
suitable stellar population model is used, some credible results can
be obtained via colours of galaxies. It is will be useful for
studying the formation and evolution of galaxies. However, some
points should be noted in studies.

\section*{Acknowledgment}
We are grateful to Profs. Zhanwen Han, Xu Kong, Licai Deng, and Gang
Zhao for useful discussions.

\label{lastpage}

\clearpage

\end{document}